\documentclass[aps,prl,twocolumn,groupedaddress,showpacs]{revtex4}

\usepackage[pdftex]{graphicx}

\usepackage[utf8]{inputenc}

\usepackage{color} 
\usepackage{hyperref}

\begin{document}

\normalsize
\title{Indirect evidence for strong coupling superconductivity in FeSe
under pressure from first-principle calculations}

\author{Christoph Heil}
\email[]{cheil@sbox.tugraz.at}
\affiliation{Institute of Theoretical and Computational Physics, University of 
Technology Graz, 8010 Graz, Austria}

\author{Markus Aichhorn}
\affiliation{Institute of Theoretical and Computational Physics, University of 
Technology Graz, 8010 Graz, Austria}

\author{Heinrich Sormann}
\affiliation{Institute of Theoretical and Computational Physics, University of 
Technology Graz, 8010 Graz, Austria}

\author{Ewald Schachinger}
\affiliation{Institute of Theoretical and Computational Physics, University of 
Technology Graz, 8010 Graz, Austria}

\author{Wolfgang von der Linden}
\affiliation{Institute of Theoretical and Computational Physics, University of 
Technology Graz, 8010 Graz, Austria}

\date{\today}

\begin{abstract}
Whether the pairing mechanism for superconductivity in
iron-based superconductors is of itinerant or local character is a question
still debated on. In order to investigate the influence of Fermi
surface nesting, we calculate from first-principles the static susceptibility of FeSe  
under applied pressures, both hydrostatic and non-hydrostatic.
We show that the electronic band structures are highly sensitive to
the way pressure is applied and confront our theoretical results with conclusions drawn from experiments.
Given that the critical temperature of FeSe is quite universal as function of pressure, this is clear
evidence that in FeSe the evolution of
Fermi surface nesting cannot
account for the evolution of the critical temperature with pressure. 
Hence we argue that the pairing in iron-chalcogenide compounds should
be of strong-coupling spin fluctuation origin. 
\end{abstract}


\maketitle

The discovery of a new class of iron based high-T$_c$ superconductors
in 2008 \cite{kamihara_iron-based_2008} has kicked off an avalanche of
research in solid state physics, a lot of it dedicated to the key
question of the 
superconducting pairing mechanism.  
A feature that has been agreed upon is 
that the parent
compounds of the FeAs superconductors exhibit an antiferromagnetic
(AFM) spin density wave (SDW) phase, as seen in
\textit{Ln}OFeAs (\textit{Ln} = La,Ce,Pr,Sm
\cite{cruz_magnetic_2008,zhao_structural_2008,drew_coexistence_2009}) 
and \textit{Ae}Fe$_2$As$_2$ (\textit{Ae}=Ba,Sr,Ca
\cite{huang_neutron-diffraction_2008,zhao_spin_2008,goko_superconducting_2009}),
or are on the verge of it, as in \textit{A}FeAs (\textit{A} = Li,Na
\cite{tapp_lifeas:_2008,chu_synthesis_2009}). 
In such materials, superconductivity (SC) arises or is enhanced, 
respectively, when this
long-range magnetic order is suppressed by doping or by application of
physical or chemical pressure. 
In this respect, pnictide and chalcogenide compounds are much more
flexible as compared to cuprate high-T$_c$ superconductors, where
the superconducting state can be reached only by doping, and intercalating layers
between the copper-oxygen planes are therefore essential.
This large flexibility has been underpinned by the discovery of
SC in FeSe \cite{hsu_superconductivity_2008,fang_superconductivity_2008,yeh_tellurium_2008,mcqueen_extreme_2009},
a material lacking completely 
the intercalated planes of the iron-pnictides.  
This structural simplicity and its enormous response to external
pressure (T$_c$ increases from $8$\,K at ambient pressure~\cite{hsu_superconductivity_2008} to $37$\,K at
about $7$\,GPa~\cite{margadonna_pressure_2009}) made it a popular testing
ground to study mechanisms of SC in
iron-based superconductors. 

Theoretical studies have shown~\cite{aichhorn_theoretical_2010,miyake_comparison_2010,yin_kinetic_2011} that 
iron chalcogenides are significantly stronger correlated than the iron
pnictide materials. Since this goes hand-in-hand with the enhancement of
localized magnetic moments, the applicability of the Fermi surface
nesting (FN) scenario in this class of compounds has to be clarified. In that scenario, SC can be 
regarded as a consequence of perturbations of the AFM ordering of the parent material~\cite{subedi_density_2008,mizuguchi_substitution_2009,medvedev_electronic_2009,kumar_crystal_2010,ciechan_pressure_2011,pan_distorted_2011}, 
and suppression of peaks in the susceptibility due to FN is an
indicator for the occurrence of SC.  

In order to shed light on this issue we perform first-principle
calculations of the static susceptibility of FeSe under 
pressure.
Many independent experimental studies on pressure vs. $T_c$ phase
diagrams have been published for
FeSe~\cite{margadonna_pressure_2009,medvedev_electronic_2009,bendele_pressure_2010,okabe_pressure-induced_2010,miyoshi_magnetic_2010}. The 
general consensus seems to be that $T_c$ starts around 8\,K for ambient 
pressure, has a maximum at about $7$\,GPa and
decreases again for higher pressures.
While all the works mentioned above applied pressure hydrostatically, a recent 
study investigated 
how T$_c$ changes under {\em non}-hydrostatical 
pressure~\cite{uhoya_simultaneous_2012}. It is interesting to observe that the 
corresponding phase diagram
is quite similar, i.e. the way how pressure is applied seems to be of minor 
importance for SC.

Our results show that suppression of FN cannot be reconciled with the
occurrence of SC. Furthermore, the way how pressure is
applied to the system substantially changes the electronic band
structure and consequently the static susceptibility. Given the universality
of $T_c$ vs. pressure, this indicates that SC in FeSe is
driven by short-ranged spin fluctuations between rather localized magnetic
moments~\cite{bao_tunable_2009,bendele_pressure_2010,kumar_pressure_2011}.

{\it Method:}
A very useful quantity for theoretical investigations of the strength
and changes of the FN is the static susceptibility matrix $\chi^{0}(\mathbf{q})$ (in short $\chi^{0}$). 
For a system of Bloch electrons - taking into account only the diagonal 
elements of this matrix - it reads in the random phase approximation

\begin{widetext}
\begin{equation}
\chi^0(\mathbf{q}) = \frac{1}{2 \pi^3} \sum \limits_{n, m} \int 
\limits_{\text{(BZ)}} d\mathbf{k} \ 
\frac
{f_m(\mathbf{k})\, [1-f_n(\mathbf{k}+\mathbf{q}+\mathbf{G})]}
{\epsilon_m(\mathbf{k})-\epsilon_n(\mathbf{k}+\mathbf{q}+\mathbf{G})} \,
|\langle m,\mathbf{k}| e^{-i\mathbf{q}\cdot\mathbf{r}} | 
n,\mathbf{k}+\mathbf{q}+\mathbf{G}\rangle|^2~,
\label{eq:chi}
\end{equation}
\end{widetext}

\noindent
where $\mathbf{q}$ represents a vector of the {\em extended} wave number space,
$n$ and $m$ denote electron band indices, $\mathbf{k}$ belongs to the first
Brillouin zone (BZ), and $\epsilon_m(\mathbf{k})$ and $f_m(\mathbf{k})$ 
mean the energy dispersion of the $m$th band and the Fermi function, 
respectively.
The reciprocal lattice vector $\mathbf{G}$ is defined such that
$\mathbf{k}+\mathbf{q}+\mathbf{G} \in$ BZ.
We want to stress that the matrix elements
\mbox{$\langle m,\mathbf{k}| e^{-i\mathbf{q}\mathbf{r}} | n,\mathbf{k}+\mathbf{q}+\mathbf{G}\rangle$}
are fully included in our calculation.

FN leads to a pronounced peak of $\chi^{0}$ at the
nesting vector $\mathbf{q} = \mathbf{q}_N$ due to
singularities of the integrand.  
Many undoped parent materials of iron-based superconductors show a marked FN and consequently a 
maximum of $\chi^{0}$ at $\mathbf{q}_N$. 
A perturbation of the crystal structure due to applied pressure or of the 
electronic band structure due to 
(electron or hole) doping is able to suppress the FN and the strong
response of $\chi^{0}$. It is argued 
\cite{monthoux_magnetically_2001,moriya_antiferromagnetic_2003,mazin_unconventional_2008,kuroki_unconventional_2008} that the reduction of the collectively 
AFM-ordered electron spins in favor of the appearance of spin-flip 
processes may act as a generator for electron pairing and SC. 

\begin{figure}[t]
  \begin{center}
     \includegraphics[width=0.8\linewidth]{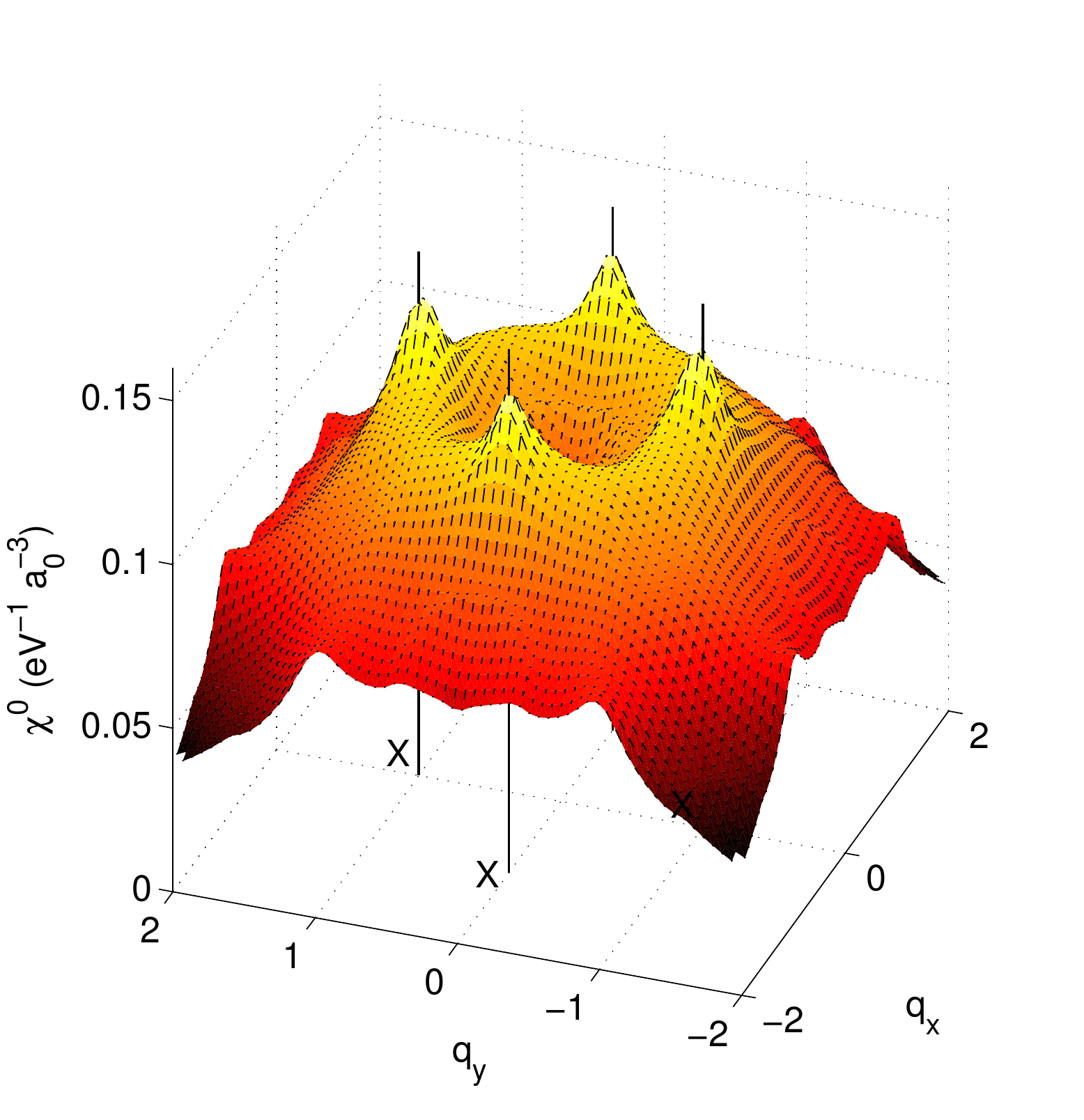}
    \caption{Static susceptibility $\chi^0$ in the $(q_x,q_y,0)$-plane
      of FeSe at a hydrostatic pressure of $2.8$\,GPa. The
      vertical solid black lines mark the positions of the X points
      in the reciprocal unit cell and $q_x$, $q_y$ are given in units $2\pi/a$ and $2\pi/b$, respectively.} 
    \label{fig:FeSe_2_8_k_cmma_surf}
  \end{center}
\end{figure}

\begin{figure}[t]
  \begin{center}
     \includegraphics[width=0.8\linewidth]{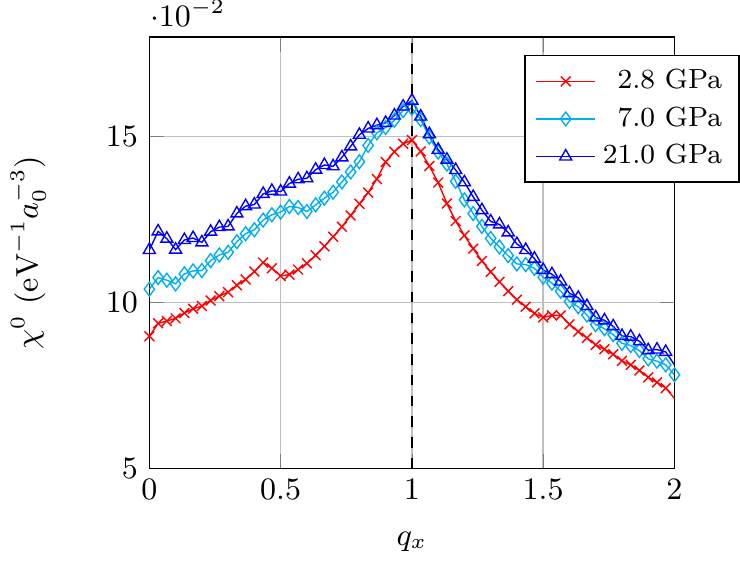}
\caption{Static susceptibility $\chi^0$ in the $(q_x,0,0)$ direction of 
FeSe for different hydrostatic pressures ($q_x$ is given in units $2\pi/a$). Apart from 
the vertical shift, due to the increasing density of states at the Fermi energy, $\chi^0$ shows very
little dependence on pressure.} 
\label{fig:FeSe_comp_rpol_total_kumar}
  \end{center}
\end{figure}

All results presented in this work have been obtained on the basis of non-spinpolarized DFT FPLAPW band structure
calculations using the WIEN2k package~\cite{blaha_wien2k_2001}. The
exchange-correlation potential has been approximated by the generalized gradient approximation GGA, 
and the base-centered orthorhombic crystal
structure Cmma has been used. The numerical integration over 
the first Brillouin zone has been performed with a semi-analytical tetrahedron
method. Each electron wave function has been represented by about 4000
plane waves, and 52 electron bands, corresponding to an energy interval 
from $-1.2$\,Ry below to $3$\,Ry above
the Fermi energy, have been included.
The momentum vectors $\mathbf{q}$ are expressed in units of 
$(2\pi/a,2 \pi/b,2 \pi/c)$ throughout this work, where $a,b$ and $c$ are the FeSe unit cell
lattice parameters, and $\chi^0$ is presented in units $(\text{eV} \cdot 
a_0^3)^{-1}$ with $a_0$ being the Bohr radius.
The lattice and unit cell parameters are taken from
experiment, where we used data by Margadonna~{\em et al.}~\cite{margadonna_pressure_2009} and Kumar~{\em et al.}~\cite{kumar_crystal_2010} for hydrostatic pressures, and for the non-hydrostatic case data from Uhoya~{\em et al.}~\cite{uhoya_simultaneous_2012}.

{\it Results:} We
start the discussion of the pressure dependence 
with the hydrostatic case. 
We checked that the results based on the crystal structures reported by Kumar~{\em et al.}~\cite{kumar_crystal_2010} are essentially the same compared to the results based on structures reported by Margadonna~{\em et al.}~\cite{margadonna_pressure_2009}. In the following, we will hence show only results based on data by Kumar~{\em et al.}. In Fig.~\ref{fig:FeSe_2_8_k_cmma_surf} a surface plot of the static susceptibility over the \mbox{$(q_x,q_y,0)$}-plane for FeSe at a
hydrostatic pressure of $2.8$\,GPa is presented. The dominating features are the 
marked peaks, indicating a strong FN of the hole 
pockets at the $\Gamma$ point and the electron pockets at the X
points, corresponding to the FN vector $\mathbf{q}_N=(1,0,0)$. For a
better quantitative discussion of the pressure dependence, we show a
$(q_x,0,0)$ cut of the susceptibility in
Fig.~\ref{fig:FeSe_comp_rpol_total_kumar}. In this diagram we compare the
susceptibility in the $(1,0,0)$ direction for three structures at pressures 
$2.8$\,GPa (red, $\times$), $7$\,GPa (cyan, $\diamond$) and
$21$\,GPa (blue, $\triangle$). Interestingly, the three calculated
profiles have a quite similar shape. They differ only by a vertical
shift, reflecting the increase of the electronic density of states
$N(\epsilon)$ at the Fermi energy $\epsilon_F$
with increasing pressure, according to the relation 
$\chi^0(0)=N(\epsilon_F)/\Omega_0$ where $\Omega_0$ denotes the volume of the 
unit cell of the crystal.
However, the prominent peak at $\mathbf{q}_N=(1,0,0)$
does not vanish with increasing pressure, leading to the
conclusion that the strength of the FN is not pressure-dependent in
the hydrostatic case.

\begin{figure}[t]
  \begin{center}
     \includegraphics[width=0.8\linewidth]{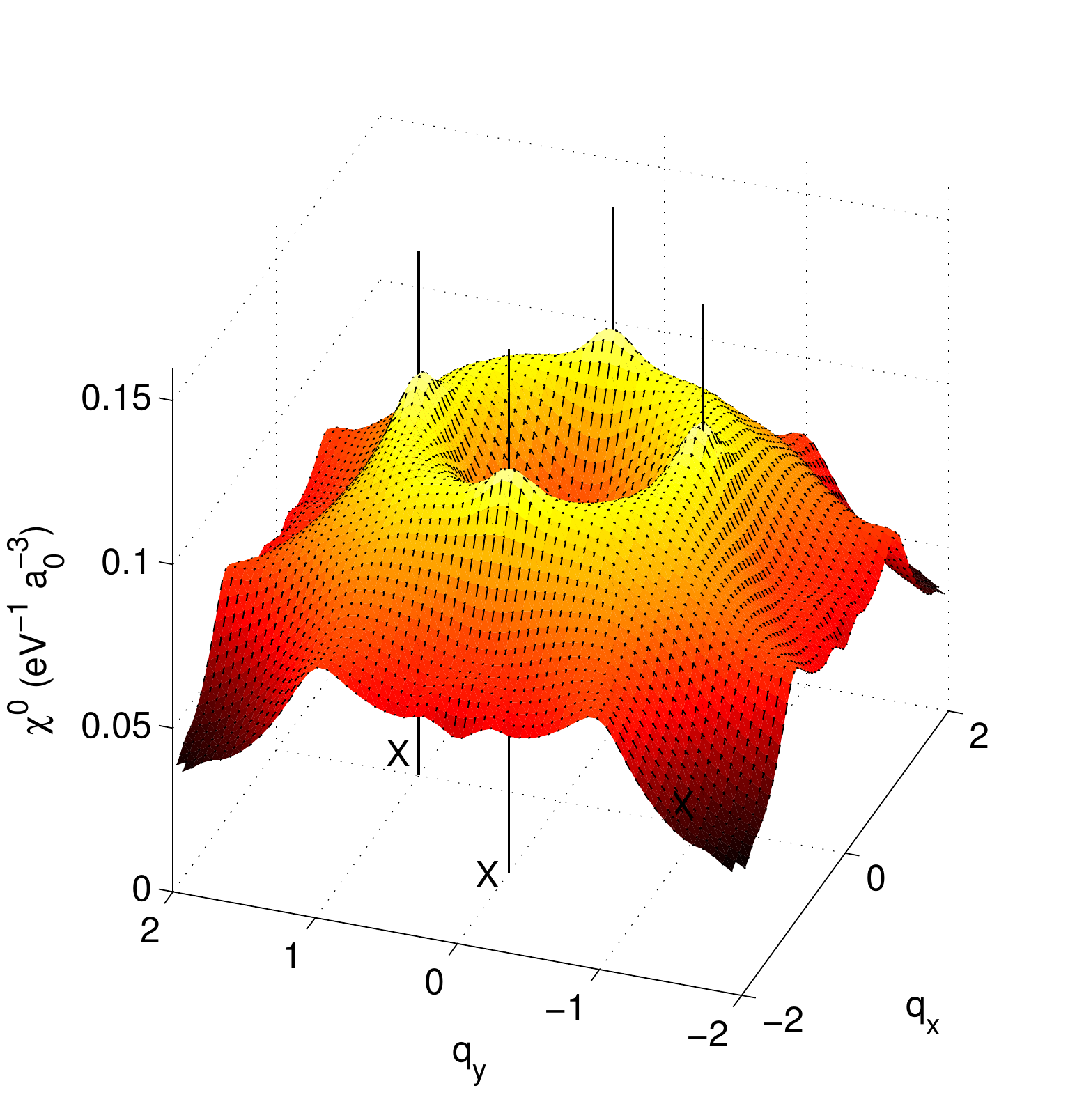} \\
    \caption{Static susceptibility $\chi^0$ in the $(q_x,q_y,0)$-plane
      for FeSe at a non-hydrostatic pressure of $4$ GPa. The
      vertical solid black lines mark the positions of the X points
      in the reciprocal unit cell and $q_x$, $q_y$ are given in units $2\pi/a$ and $2\pi/b$, respectively.} 
    \label{fig:FeSe_4_u_cmma_surf}
  \end{center}
\end{figure}

\begin{figure}[t]
  \begin{center}
     \includegraphics[width=0.8\linewidth]{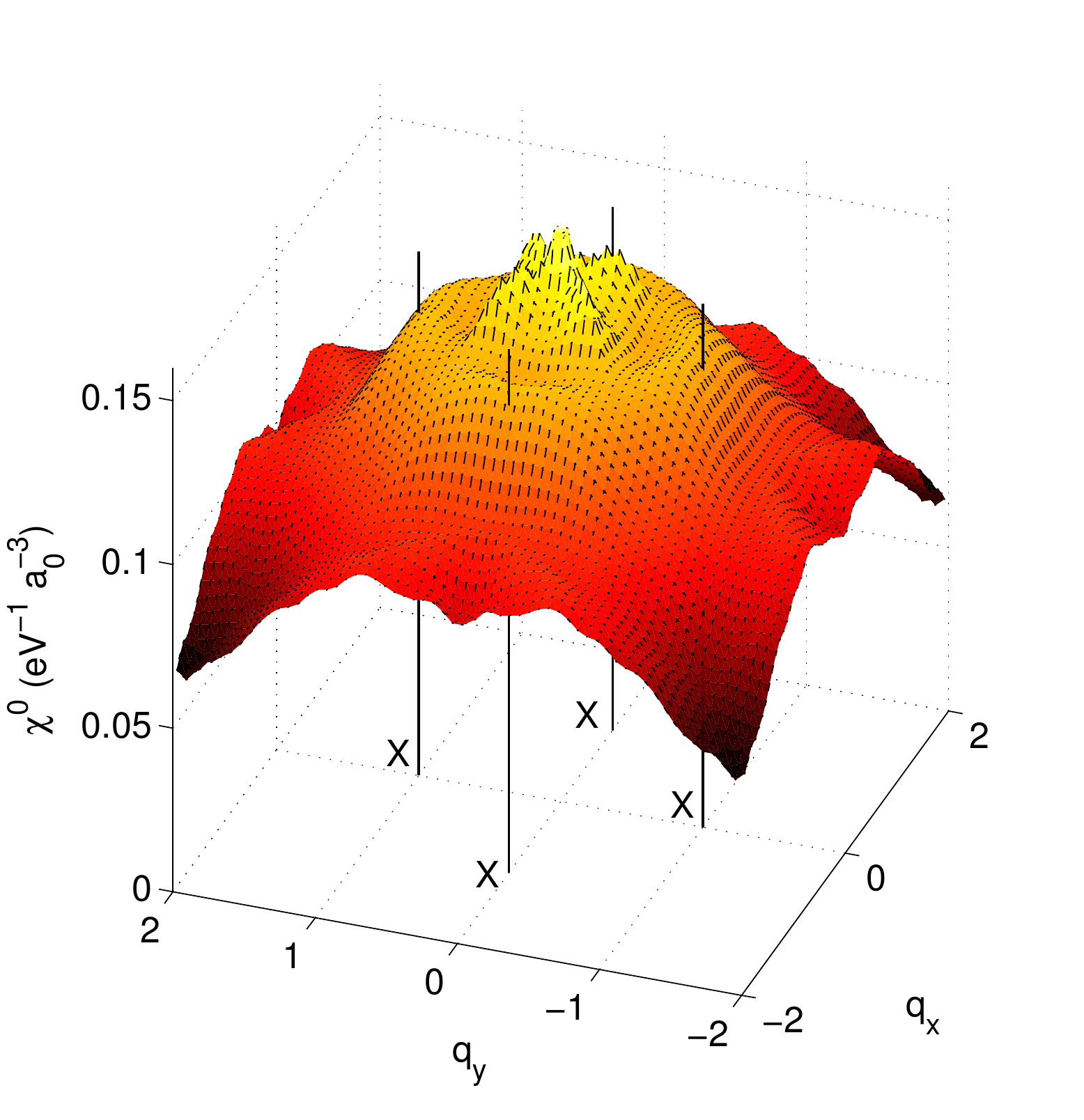} \\
    \caption{Static susceptibility $\chi^0$ in the $(q_x,q_y,0)$-plane
      for FeSe at a non-hydrostatic pressure of $12.9$
      GPa. The vertical solid black lines mark the positions of the
      X points in the reciprocal unit cell and $q_x$, $q_y$ are given in units $2\pi/a$ and $2\pi/b$, respectively.} 
    \label{fig:FeSe_12_9_u_cmma_surf}
  \end{center}
\end{figure}

\begin{figure}[t]
  \begin{center}
     \includegraphics[width=0.8\linewidth]{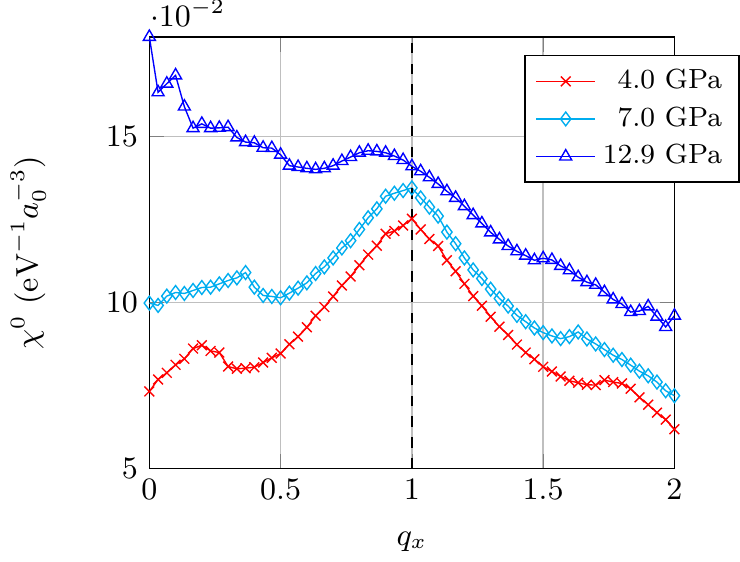}
\caption{Static susceptibility $\chi^0$ in the $(q_x,0,0)$ direction for 
FeSe for different non-hydrostatic pressures ($q_x$ is given in units $2\pi/a$). $\chi^0$ shows a marked peak at the FN vector $\mathbf{q}_N$ for the $4$\,GPa case, which is still present, albeit reduced, for the $7$\,GPa compound. At high pressures of $12.9$\,GPa no prominent feature can be seen.} 
\label{fig:FeSe_comp_rpol_total_uhoya}
  \end{center}
\end{figure}

The situation is different in the non-hydrostatic case, as demonstrated in
Figs.~\ref{fig:FeSe_4_u_cmma_surf} and
\ref{fig:FeSe_12_9_u_cmma_surf}. Here we present the static
susceptibility for the compounds reported by Uhoya~{\em et al.}~\cite{uhoya_simultaneous_2012}. Fig.~\ref{fig:FeSe_4_u_cmma_surf} depicts
the situation for $4$\,GPa, where the marked maxima at the X
points, similar to the hydrostatic case, can still be observed. On the
other hand, for the high pressure case ($12.9$\,GPa) shown in
Fig.~\ref{fig:FeSe_12_9_u_cmma_surf}, these maxima are 
reduced significantly. 
This change of the susceptibility is even better demonstrated in
Fig.~\ref{fig:FeSe_comp_rpol_total_uhoya}, where we depict  
again a $(q_x,0,0)$ cut of the susceptibility. Here we compare
$\chi^0$ for the case of non-hydrostatic pressures at $4$\,GPa (red,
$\times$), $7$\,GPa (cyan,$\diamond$) and $12.9$\,GPa
(blue,$\triangle$). For the intermediate
pressure regime of $7$\,GPa we interpolated the structural data from
Uhoya~{\em et al.}~\cite{uhoya_simultaneous_2012} for $4$\,GPa and
$12.9$\,GPa. 
It is known from experiments~\cite{margadonna_pressure_2009,kumar_crystal_2010,matsuishi_structural_2012} that - in such moderate pressure regimes - the unit cell parameters change linearly
with pressure, supporting this approach.
While the strong response at $\mathbf{q}_N$ is still present for the cases with
$4$ and $7$\,GPa (albeit slightly reduced for the $7$\,GPa structure), it has
vanished completely for the high pressure case. This behavior 
of $\chi^0$ allows the conclusion that the FN in FeSe is rather strong
for low and reduced considerably for higher non-hydrostatic pressures.

\begin{figure}[t]
  \begin{center}
     \includegraphics[width=0.7\linewidth]{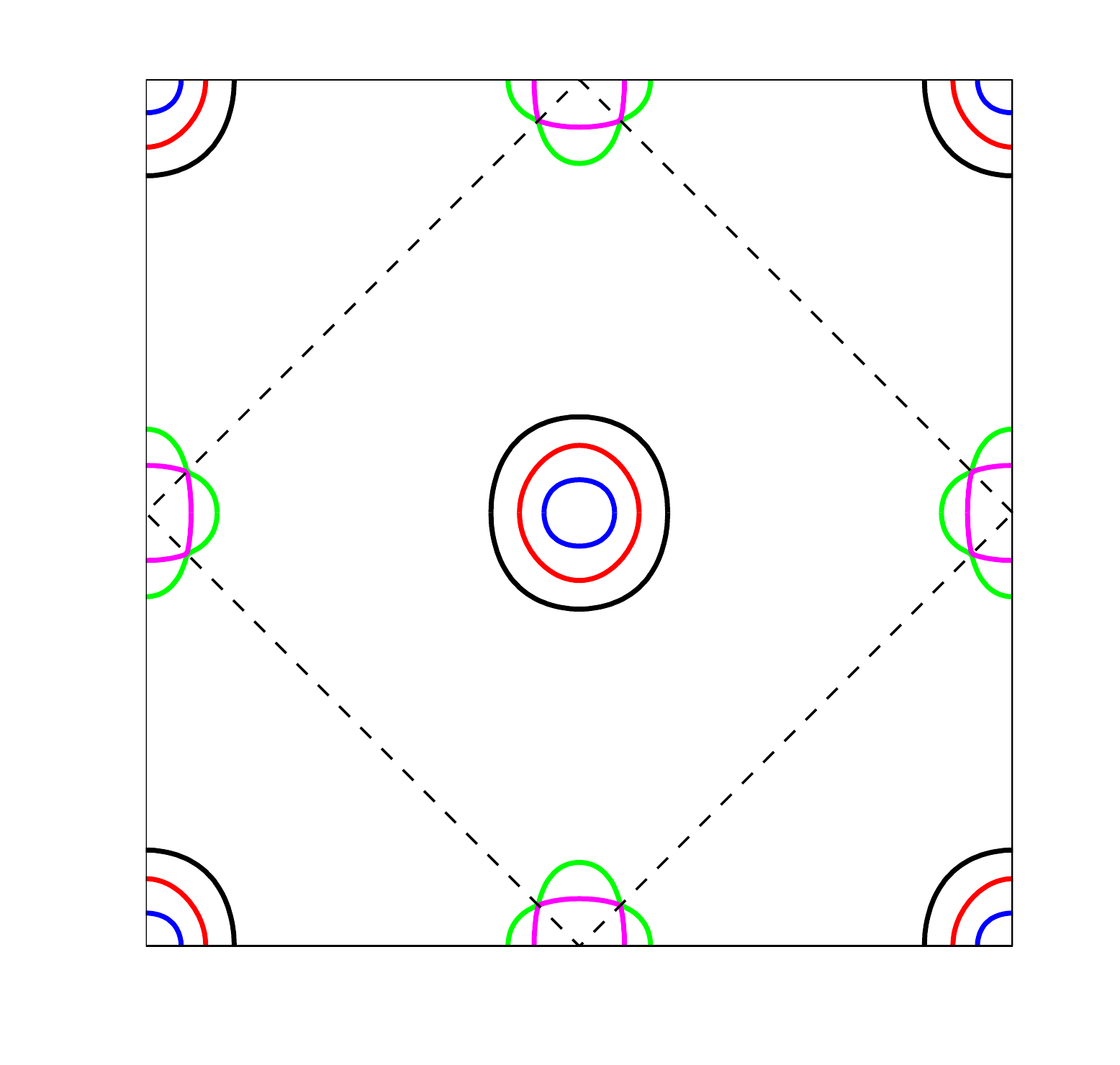}
    \caption{Fermi surface $q_z=0$ cut for FeSe at a non-hydrostatic pressure of $4$ GPa. The different colors correspond to
      different electron bands that cross the Fermi energy and the
      black dashed lines mark the boundaries of the reciprocal unit cell.}
    \label{fig:FeSe_4_u_fs_cut}
  \end{center}
\end{figure}

\begin{figure}[t]
  \begin{center}
     \includegraphics[width=0.7\linewidth]{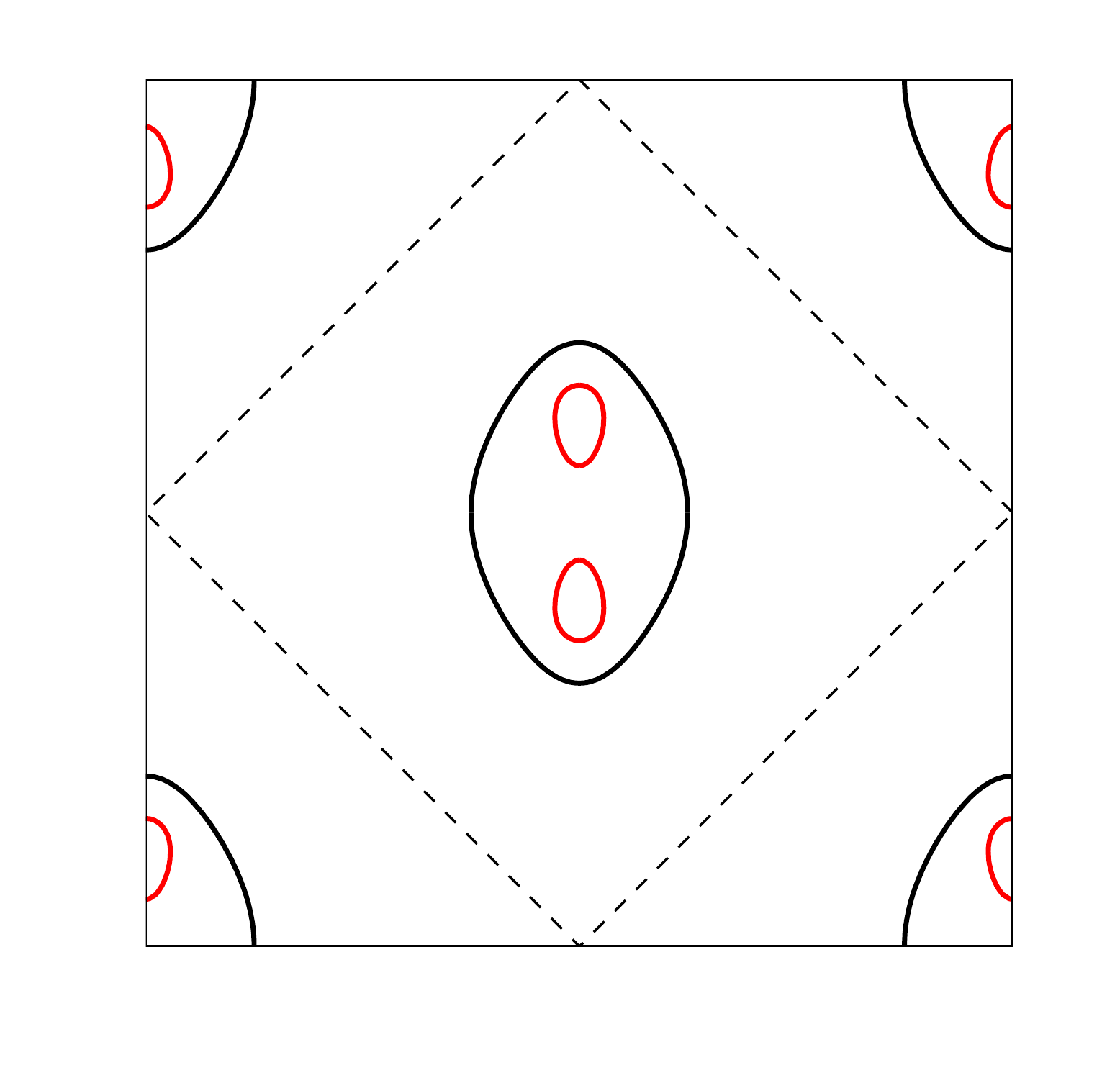}
    \caption{Fermi surface $q_z=0$ cut for FeSe at a non-hydrostatic pressure of $12.9$ GPa. The different colors correspond to
      different electron bands that cross the Fermi energy and the
      black dashed lines mark the boundaries of the reciprocal unit cell.} 
    \label{fig:FeSe_12_9_u_fs_cut}
  \end{center}
\end{figure}

In order to get better insight into this different FN behavior, we present 
in Figs.~\ref{fig:FeSe_4_u_fs_cut} and \ref{fig:FeSe_12_9_u_fs_cut} the
cuts of the Fermi surfaces in the \mbox{$(q_x,q_y,0)$}-plane for low ($4$\,GPa) and high ($12.9$\,GPa)
non-hydrostatic pressures. The different colors in
these diagrams correspond to different bands, while the black dotted
lines mark the edges of the unit cell in reciprocal space.

For low pressures (Fig.~\ref{fig:FeSe_4_u_fs_cut}),
it can clearly be seen that there is a marked FN, originating mainly from the interband transitions between the hole-like pockets around the $\Gamma$ point and the electron-like pockets around the X point.
Different to doping, where hole pockets shrink and electron
pockets grow - or vice versa - the effect of non-hydrostatic
pressures is a bit more subtle:
Large pressures lead to considerable
changes in the $a/b$ lattice parameters, and as a consequence,
not only the sizes of the pockets, but also their shapes are
altered substantially~(Fig.~\ref{fig:FeSe_12_9_u_fs_cut}).                                                                                                                                                                                                                                                                                                                                                                                                                                                                                               
In the case of low pressures the electron and hole pockets are in good approximation circularly shaped. For high pressures on the other hand, the pockets at the X points disappear completely and the Fermi surfaces at the $\Gamma$ point are distorted to more complicated elliptical shapes. This results in a complete loss of the FN vector $\mathbf{q}_N$ and further supports our conclusion drawn from Figs.~\ref{fig:FeSe_4_u_cmma_surf}-\ref{fig:FeSe_comp_rpol_total_uhoya}, namely, that an increasing non-hydrostatic pressure causes a significant decrease of the FN:
While in the $4$\,GPa case the nesting vector \mbox{$\mathbf{q}_N=(1,0,0)$} can
clearly be observed, such a feature is not visible for a pressure of $12.9$\,GPa.
We want to stress that the 
calculation of the complete susceptibility offers more detailed and reliable information
about the position of the SDW-vector and the strength of the FN. This
is based on the fact that $\chi^0$ involves the integration over the
whole Brillouin zone and the inclusion of the matrix elements, as can
be seen in Eq.~\ref{eq:chi}.

{\it Summary:} The results presented above have important implications concerning the behavior of SC in FeSe. While for low pressures the static susceptibility of this material exhibits a marked FN, the higher pressure properties depend crucially on whether pressure is applied hydrostatically or not. According to our calculations there is no visible reduction of the FN for applied hydrostatic pressures, even as large as $21$\,GPa. On the other hand, we observe that the FN is reduced significantly if pressure is applied non-hydrostatically. Based on these theoretical results alone one might argue that - in the framework of FN-mediated pairing - SC in FeSe is only supported efficiently by non-hydrostatic pressures.
However, such a result contrasts sharply with experiments, which report a considerable increase of T$_c$ with increasing pressure {\em independent} of the way pressure is applied.

This leads to the conclusion, that SC in the FeSe compounds is not based on a weak-coupling nesting between electron and hole pockets, and therefore not mediated by spin fluctuations coming from an itinerant SDW state. Moreover, recent studies have shown~\cite{ksenofontov_density_2010,luo_quasiparticle_2012,wang_electronic_2012} that the pronounced increase of T$_c$ with pressure cannot be explained by phononic Cooper-pairing, too. Therefore, staying in the paradigm of spin-mediated pairing, we would argue that the SC in FeSe is based on fluctuations between {\em more localized} magnetic moments.


\end{document}